\documentclass{article}
\usepackage{arxiv}

\usepackage[utf8]{inputenc} 
\usepackage[T1]{fontenc}    
\usepackage{hyperref}       
\usepackage{url}            
\usepackage{booktabs}       
\usepackage{amsfonts}       
\usepackage{nicefrac}       
\usepackage{microtype}      
\usepackage{lipsum}		
\usepackage{graphicx}
\usepackage[round]{natbib}
\usepackage{doi}

\usepackage{latexsym}
\usepackage{amsmath}
\usepackage{amsfonts}
\usepackage{amssymb}
\usepackage{psfrag}
\usepackage{graphicx}
\usepackage{bm}
\usepackage{makecell}
\usepackage{footmisc}
\usepackage{float}
\usepackage{siunitx}
\usepackage{verbatim} 
\usepackage{enumitem}
\usepackage{booktabs}

\usepackage{array}
\newcolumntype{C}[1]{>{\centering\arraybackslash}m{#1}}
\newcolumntype{L}[1]{m{#1}}
\usepackage{multirow}
\usepackage{bm}

\usepackage{setspace}
\title{Dynamic estimation with random forests for discrete-time survival data}

\author{Hoora Moradian \\
	Department of Decision Sciences \\
	HEC Montr\'{e}al\\
    Montr\'{e}al, Qu\'{e}bec, Canada H3T 2A7\\
	\texttt{hoora.moradian@hec.ca} \\
	\And
	Weichi Yao \\
	Department of Technology, Operations, and Statistics \\
	Stern School of Business\\
	New York University \\
	New York, NY 10012, USA\\
	\texttt{wyao@stern.nyu.edu} \\
	\And
	Denis Larocque \\
	Department of Decision Sciences \\
	HEC Montr\'{e}al\\
    Montr\'{e}al, Qu\'{e}bec, Canada H3T 2A7\\
	\texttt{denis.larocque@hec.ca} \\
	\And
	Jeffrey S. Simonoff\\
	Department of Technology, Operations, and Statistics \\
	Stern School of Business\\
	New York University \\
	New York, NY 10012, USA\\
	\texttt{jsimonof@stern.nyu.edu} \\
	\And
	Halina Frydman\\
	Department of Technology, Operations, and Statistics \\
	Stern School of Business\\
	New York University \\
	New York, NY 10012, USA\\
	\texttt{hfrydman@stern.nyu.edu } \\
}

\date{}

\renewcommand{\headeright}{Research article}
\renewcommand{\undertitle}{Research article}
\renewcommand{\shorttitle}{MORADIAN, YAO, LAROCQUE, SIMONOFF and FRYDMAN}

\begin{document}
\maketitle

\begin{abstract}
	Time-varying covariates are often available in survival studies and estimation of the hazard function needs to be updated as new information becomes available. In this paper, we investigate several different easy-to-implement ways that random forests can be used for dynamic estimation of the survival or hazard function from discrete-time survival data. The results from a simulation study indicate that all methods can perform well, and that none dominates the others. In general, situations that are more difficult from an estimation point of view (such as weaker signals and less data) favour a global fit, pooling over all time points, while situations that are easier from an estimation point of view (such as stronger signals and more data) favor local fits. 
\end{abstract}


\section{Introduction}
Survival analysis studies with time-to-event data have applications in many research areas. 
It is common in practice that the actual time until the occurrence of an event of interest is observed only for some of the subjects and only partial information about the time is available for other subjects, for example, because the study ended before all subjects experienced the event, or because some of them were lost during the study. This concept is known as censoring \citep{KleinandMoeschberger2003}. 
Right-censoring, when only a lower bound on the actual time is observed, is the most common situation and will be the main focus of this paper. 
A comprehensive introduction to modeling time-to-event data can be found in \citet{KleinbaumandKlein2005} and \citet{HosmerJretal2011}. 

Many of the traditional methods for analysing continuous time-to-event data rely on some parametric (e.g. Weibull) or semi-parametric (e.g. Cox) assumptions about the link between the covariates and the time response, which may result in poor performance in real-world applications. 
Recently, more flexible models and adapted machine learning algorithms that use data to find relevant structures, instead of imposing them a priori, have been developed in the survival analysis domain \citep{Wangetal2019}. One class of such models is tree-based methods, which are the focus of this paper. 

Tree-based methods were first developed for a categorical or continuous outcome. \citet{Breimanetal1984} is the earliest monograph about trees and details the Classification and Regression Tree (CART) paradigm. \citet{GordonandOlshen1985} extended the tree paradigm to survival data and introduced survival trees \citep{LeblancandCrowley1993, Segal1988}. However, it is well-known that ensembles of trees often provide better estimation performance than a single tree. One popular and efficient ensemble method is the random forest, introduced by \citet{Breiman2001}, and extended to model right-censored survival data \citep{Ishwaranetal2004, Hothornetal2006, Ishwaranetal2008, ZhuandKosorok2012}. 
There is a vast literature on survival trees and forests and \citet{Bou-Hamadetal2011b} present a general overview.

In many studies, an estimate of the hazard function for a subject is obtained at time 0 using only the baseline covariate information. However, when time-varying covariates are present, it is often preferable to update the estimates of hazard probabilities as new longitudinal information becomes available. This is the topic of ``dynamic estimation,'' which is a growing area of interest. There are primarily three approaches to build dynamic estimates in this context: 1) landmark analysis, 2) joint modeling, and 3) a counting process approach. The idea of landmark analysis \citep{Andersonetal1983,Madsenetal1983} is to build models, usually Cox, at different landmark times $t$ using the covariate information available up to $t$ from those subjects who are still at risk of experiencing the event at $t$. Comprehensive treatments of this approach are given in \citet{vanHouwelingen2007} and \citet{vanHouwelingenandPutter2011}. The second approach uses joint modeling of the time-varying covariates processes and the event time data process \citep{Hendersonetal2000}. This approach depends on the correct specification of the model for the time-varying covariates trajectories, and this problem amplifies as the number of time-varying covariates increases. The main idea of the third approach is to partition the followup information for each individual into multiple segments on non-overlapping intervals \citep{BacchettiandSegal1995}. This is used to accommodate time-varying covariates in the tree building process \citep{Bertoletetal2016, FuandSimonoff2017a}. Survival forest algorithms based on this same counting process approach can then be developed to provide dynamic estimation of hazards or survival probabilities \citep{Wongvibulsinetal2020,Yaoetal2020}.

Most of the research, including the work presented above, assumes that the time-to-event is measured continuously, when in fact it is measured on a discrete scale in many cases. This can happen with binned data where the event occurs in an interval of time, which are not necessarily of the same length. 
For example, the Framingham Heart Study \footnote{https://www.nhlbi.nih.gov/science/framingham-heart-study-fhs} requires the participants to return to the study approximately every two to six years in order for their medical history data to be collected and physical exams and laboratory tests done. 
Another example of binned data is term insurance, or any other annual contract with churn (lack of renewal of the contract) being the event of interest.
Alternatively, the observed time may come from a truly discrete process, such as the number of elapsed time units or trials before reaching a specific goal (e.g. the number of cycles until pregnancy). 
Although traditional modeling approaches for continuous-time survival data can also be applied to discrete-time survival data, \citet{TutzandSchmid2016} explain the advantages of using statistical methods that are specifically designed for discrete event times.  
They point out that the hazard functions derived in the discrete case are more easily interpretable than for continuous survival time data, since the hazards can then be formulated as conditional probabilities. Moreover, discrete models do not have any problems dealing with ties. Therefore, in this paper, we only focus on the methods specifically designed for discrete-time survival data.

Survival trees and forests designed specifically for discrete-time responses were developed by \citet{Bou-Hamadetal2009}, \citet{Bou-Hamadetal2011a}, \citet{Schmidetal2016}, \citet{Bergeretal2019} and \citet{Schmidetal2020}. 
Section \ref{sec:description_existing} provides a description of some of these methods since they are central to this article. \citet{Elgmatietal2015} propose a penalized Aalen additive model for dynamic estimation of the hazard function for discrete-time recurrent event data, but the method is limited to one-step ahead estimation, while we also explore multi-step ahead estimation.

From the above discussion, we see that no tree-based methods have addressed the problem of dynamic estimation with discrete survival responses. In this paper we investigate different ways that random forests can be used for dynamic estimation of hazard function with discrete-time survival response data. 

The rest of the paper is organized as follows. Section \ref{sec:description} describes the data setting and the proposed methods. The results from a simulation study are presented in Section \ref{sec:simulations}.  
Section \ref{sec:conclusions} provides conclusions and directions for future work. More details about the simulation study and a real data example using bankruptcy data can be found in a separate supplemental material document.

\section{Description of the methods} \label{sec:description}
Suppose we have data on $N$ independent subjects. For subject $i$, observations are in the form of ($\tau_i$, $\delta_i$, $\bm{x}_i$) where $\tau_i \in \{{1,2,\ldots,T}\}$ is the discrete time to event, $T$ is the maximum observed time in the data set, $\delta_i$ is the censoring index which takes a value of 0 if the observation for the subject $i$ is right censored and a value of 1 if subject $i$ has experienced the event of interest, and $\bm{x}_i$ is a set of covariates, some of which can be time-varying and some time-invariant. We will denote by $x_{ki}(t)$ the value of the $k^{th}$ covariate, $k \in \{1,2,\ldots,p\}$, at time $t \in \{0,1,\ldots,T-1\}$ for subject $i$. Hence, $\bm{x}_i(0)$ gives the baseline covariate values. For simplicity, we will use this notation for all covariates, time-varying or not.
Hence $x_{ki}(t)$ remains constant for all $t$ for a time-invariant covariate.
The values of the actual time to event and the censoring times for subject $i$ are denoted by $U_i$ and $V_i$, respectively. Hence we have $\tau_i=\min(U_i,V_i)$ and we assume that $U_i$ and $V_i$ are independent given $\bm{x}_i$. The hazard function for subject $i$ is denoted by $h_i(t) = P(U_i = t \mid U_i \geq t)$ for simplicity but it is obvious that $\tau_i$, $\delta_i$, $U_i$ and $V_i$ depend on $\bm{x}_i$.
Similarly, the survival function for subject $i$ is $S_i(t)=P(U_i>t)$, and the probability that the event occurs at time $t$ is $\pi_i(t)=P(U_i=t)$.
These two functions can be obtained from the hazard function with the recursive formulae $S_i(t)=S_i(t-1)(1-h_i(t))$ and $\pi_i(t)=S_i(t-1)-S_i(t)$, with $S_i(0)=1$.
Hence, it is sufficient to model the hazard function (or any one of the other two functions) to recover the other ones.

\subsection{Description of existing methods for discrete-time survival data}\label{sec:description_existing}
The existing methods for dynamic estimation based on time-varying covariates data rely on the counting process approach to reformat the data. To fix ideas, a generic data set of ten observations with two covariates, $X_1$ being time-varying and $X_2$ being time-invariant, is given in Table \ref{generic}. For instance, the first subject experienced the event at the second time point and thus values of the time-varying covariate $X_1(t)$ are only available up to $t=1$, with NA's for the other time points. 
Note that we do not assume that the covariate values at the event or censoring time are available (e.g. the event or censoring may occur before the observation of the covariates). This process is repeated for each of the subjects in the data set. The reformatted data set is often called the ``person-period'' data set.

\begin{table}[t]
\centering
\caption{A generic data set with ten observations and two covariates, with $X_1$ being time-varying and $X_2$ time-invariant.}
\begin{tabular}{llllL{1cm}L{1cm}L{1cm}L{1cm}L{1cm}L{1cm}}
\toprule 
id & $\tau$ & $\delta$ & $X_1(0)$ & $X_1(1)$ & $X_1(2)$ & $X_1(3)$ & $X_1(4)$ & $X_2$ \\ 
\midrule 
\midrule 
1  & 2  & 1  & $x_{11}(0)$  & $x_{11}(1)$  & NA           & NA           & NA   & $x_{21}$  \\ 
\midrule 
2  & 4  & 1  & $x_{12}(0)$  & $x_{12}(1)$  & $x_{12}(2)$  & $x_{12}(3)$  & NA   & $x_{22}$  \\ 
\midrule 
3  & 3  & 0  & $x_{13}(0)$  & $x_{13}(1)$  & $x_{13}(2)$  & NA           & NA   & $x_{23}$  \\ 
\midrule 
4  & 1  & 0  & $x_{14}(0)$  & NA           & NA           & NA           & NA   & $x_{24}$  \\ 
\midrule 
5  & 4  & 1  & $x_{15}(0)$  & $x_{15}(1)$  & $x_{15}(2)$  & $x_{15}(3)$  & NA   & $x_{25}$  \\ 
\midrule 
6  & 4  & 0  & $x_{16}(0)$  & $x_{16}(1)$  & $x_{16}(2)$  & $x_{16}(3)$  & NA   & $x_{26}$  \\ 
\midrule 
7  & 2  & 0  & $x_{17}(0)$  & $x_{17}(1)$  & NA           & NA           & NA   & $x_{27}$  \\ 
\midrule 
8  & 4  & 1  & $x_{18}(0)$  & $x_{18}(1)$  & $x_{18}(2)$  & $x_{18}(3)$  & NA   & $x_{28}$  \\ 
\midrule 
9  & 3  & 1  & $x_{19}(0)$  & $x_{19}(1)$  & $x_{19}(2)$  & NA           & NA   & $x_{29}$  \\ 
\midrule 
10 & 4  & 0  & $x_{110}(0)$ & $x_{110}(1)$ & $x_{110}(2)$ & $x_{110}(3)$ & NA   & $x_{210}$ \\ 
\bottomrule
\end{tabular}
\label{generic}
\end{table}

We describe first the existing approaches for estimating the hazard of a subject at the $u$-th discrete time point that use the last available values of the time-varying covariates. One widely-used method is the discrete-time proportional odds (DTPO) model, which is also known as the continuation ratio model
\begin{equation}
\label{eq1}
\begin{split}
    \log \left( \frac{h_i(u)}{1-h_i(u)} \right)  =& \;\alpha_1 D_{1i}(u)+\cdots+\alpha_T D_{Ti}(u) \\ 
    & + \beta_{1} X_{1i}(u-1)+\cdots+\beta_{p}
    X_{pi}(u-1), \\
    \end{split}
\end{equation}
for $i=1,2,\ldots, n$ and $u=1,2,\ldots, T$, where the $D_{ri}(u)$'s are indicator variables indexing the $r^{th}$ discrete time point that are defined by $D_{ri}(u)=1$ if $r=u$ and 0 otherwise. The intercept parameters $\alpha_1,\alpha_2,\ldots,\alpha_T$ define the baseline hazard at each time point and the $\beta$ coefficients describe the effects of covariates on the baseline hazard function. Applying the counting process approach to reformat the generic data set gives the person-period data in Table \ref{ppdata}. The model parameters in (\ref{eq1}) can then be estimated by fitting a logistic regression to the reformatted data set (more detail can be found in \citet{WillettandSinger1993}, p. 171).

\begin{table}[t]
\centering
\caption{Person-period data set using counting process approach for DTPO model. Only the first two subjects (up to id $=2$) are shown to save space. It has one row of observation for each discrete time point $u$ in which the subject is at risk of experiencing the event and the response $y$ equals 1 if the event occurred at that time and 0 otherwise.}

\begin{tabular}{ccccccccccc}
\toprule 
id          & $y$        & $u$ & $D_1$ & $D_2$ & $D_3$ & $D_4$& $X_1$ & $X_2$ \\ 
\midrule 
\midrule 
1           & 0          & 1   & 1 & 0 & 0 & 0    & $x_{11}(0)$      & $x_{21}$         \\ 
\midrule 
1           & 1          & 2   & 0 & 1 & 0 & 0    & $x_{11}(1)$      & $x_{21}$         \\ 
\midrule 
2           & 0          & 1   & 1 & 0 & 0 & 0      & $x_{12}(0)$      & $x_{22}$         \\ 
\midrule 
2           & 0          & 2   & 0 & 1 & 0 & 0     & $x_{12}(1)$      & $x_{22}$         \\ 
\midrule 
2           & 0          & 3   & 0 & 0 & 1 & 0      & $x_{12}(2)$      & $x_{22}$         \\ 
\midrule 
2           & 1         & 4   & 0 & 0 & 0 & 1     & $x_{12}(3)$      & $x_{22}$         \\ 
\bottomrule
\end{tabular}
\label{ppdata}
\end{table}

\citet{Bou-Hamadetal2011a} were the first to propose building trees and forests using the person-period data set with $y$ as the response and a likelihood-based splitting criterion. 
\citet{Schmidetal2016} propose a classification tree by applying the CART algorithm based on the Gini impurity measure \citep{Breimanetal1984} to the same data set again with $y$ as the response. \citet{Schmidetal2020} propose building discrete-time random survival forests using Hellinger's distance criterion \citep{Cieslaketal2012} as the splitting rule. The Hellinger's distance criterion is also implemented in a classification tree approach for the modeling of competing risks in discrete time \citep{Bergeretal2019}. Numerical results given in \citet{Schmidetal2020} suggest that node splitting by the Hellinger's distance improves the performance when compared to skew-sensitive split criteria such as the Gini impurity. This is consistent with the results of simulations performed here, and therefore we only investigate forest methods using the Hellinger's distance criterion. The time point $u$ itself is also included as an ordinal covariate \citep{Schmidetal2016, Bergeretal2019, Schmidetal2020}. To fix ideas, with the data set in Table \ref{ppdata}, this means building a classification forest with $y$ as the response using the three covariates $X_1$, $X_2$ and the time point $u$. 
Using the time point as a predictor implies that the subjects can be split apart in the person-period data, even if no time-varying covariates are present among the original covariates, since the time point itself is a time-varying covariate. In a terminal node, the estimate of the hazard is the proportion of $1$ (events) in the node.

\subsection{Description of the setup for dynamic estimation}\label{sec:description_investigated}
In line with the purpose of dynamic estimation, where we want to estimate future risks, at the current time point $t$, the goal is to estimate the hazard of a subject at some future time point $u$ for $u=t+1,t+2, \ldots, T$. We assume that measurements for all covariates are available at $0, 1,2,\ldots,t$, and the methods are entitled to use all of that information. Hence, all covariate information up to time $t$ can be used to estimate the hazard function at $u$. Table \ref{tbl:problems} illustrates the possible combinations of $t$ and $u$ with $T=4$ as an example. One can also see that, for a given value of $T$, the total number of possible estimation problems is $T(T+1)/2$ ($=10$ when $T=4$). For the following discussion, $t$ always denotes the current time point, $u$ always denotes the future time point we are interested in for estimation, and $u>t$ by definition.

\begin{table}[t]
\centering
\caption {The $10$ different estimating problems when $T=4$. For instance, at time point $t=2$, given a subject who has survived up to this time point, we are interested in estimating its hazard function at the future time points $u=3,4$.}
\begin{tabular}{ C{1cm} C{0.5cm}  C{0.5cm} C{0.5cm} C{0.5cm} }
\toprule
t & \multicolumn{4}{c}{u} \\
\midrule 
Value  & 1 & 2 & 3 & 4 \\
\midrule 
0 & $\checkmark$	& $\checkmark$	& $\checkmark$	& $\checkmark$ \\
\midrule 
1 &	& $\checkmark$	& $\checkmark$	& $\checkmark$ \\
\midrule 
2 & & & $\checkmark$	& $\checkmark$ \\
\midrule 
3 & & &	& $\checkmark$ \\
\bottomrule

\end{tabular}
\label{tbl:problems}
\end{table}

For simplicity of the presentation, we will only use the last available value of the time-varying covariates to build the models. However, without loss of generality, we can assume that any past information we also want to use is already incorporated into the covariates at the current time point $t$. For example if we want to use the lag of a time-varying covariate, say $X_1(t-1)$, we can simply define a new covariate at time $t$ to represent this lag, that is, $\Tilde{X}_1(t)=X_1(t-1)$.

We investigate different methods to solve the hazard function estimation problem for each pair of $(t,u)$ as illustrated in Table \ref{tbl:problems}. These methods can be divided into three main approaches to address the same given estimation problem based on how they make use of the information provided in the generic data set, i.e. how they construct the training data sets. 

Given the estimation problem for a specific pair $(t^\ast,u^\ast)$, the first approach is to only use corresponding local information to train the model. 
More precisely, to construct the training data set to estimate the hazard for the given pair $(t^\ast,u^\ast)$, we consider only the subjects that are still alive and not censored at time point $u^\ast-1$, in order that these subjects are still at risk of experiencing the event at time point $u^\ast$. Moreover, the training data set only contains their covariate information at the current time point $t^\ast$. 
For a subject with covariate information available up to time $t^\ast$, this approach builds separate models to estimate the hazard function at each future time point.
Using separate models might be effective if the hazards at different time points are related to different covariate patterns, but this approach will likely lose efficiency when the hazards are related to similar covariate patterns because of the variability induced by using separate models.

The second approach solves the estimation problems for all future time points at a given time point $t^\ast$ at once. In this case, for a given $t^\ast$, we construct a single training data set that pools the local information $(t^\ast,u)$ from all possible values of $u$, which can reduce the variability when the hazards at a given time point are related to similar covariate patterns. 
All of the covariates are used and the future time point $u$ itself is also considered as a covariate. The model trained with this data set is then used to estimate all future hazards for any subject with its current covariate information at the given time $t^\ast$.
The \citet{Schmidetal2020} method builds the forest based on this idea and was presented in the last section.

The third approach is inspired by the so-called ``supermodel'' based on stacked data used in landmark analysis, presented by \citet{vanHouwelingen2007} and \citet{vanHouwelingenandPutter2011}.
Instead of pooling the information from the different estimation horizons only for a given $t^\ast$, as in the second approach, we can go a step further and pool all of the information for all combinations of $(t,u)$ together. The idea is to borrow information from different values of $t$, in addition to that of different estimation horizons for a given $t^\ast$.
This results in a super person-period training data set that is created by stacking the training data sets from all values of $t$ used in the \citet{Schmidetal2020} method described above. 
The model trained on this super person-period data set is then used to estimate hazard probabilities for a subject at any future time points with covariate information available at any current time point.
This time, both the estimation horizon $u$ and the value of $t$ are potential covariates, in addition to the other covariates.

\begin{table}[!t]
    \centering
    \caption{Training data set used for the three approaches to solve each of the estimation problems given in Table \ref{tbl:problems} ($T=4$): i) the first approach -- separate; ii) the second approach -- the \citet{Schmidetal2020} method; iii) the third approach -- super person-period. Only the first three subjects (up to id $=3$) are shown to save space. }

    \begin{tabular}{|lll|cccc|c|c|c|}
        \hline
      &      &      & \multicolumn{4}{c|}{\multirow{2}{*}{Available covariates}} & \multicolumn{3}{c|}{Box of data used to train a given Method (Covariates used)}\\
      &      &      &                  &            &        &        & \multicolumn{3}{c|}{ to estimate hazards for which value of $(t,u)$.}\\
      \cline{4-10} 
    \multirow{2}{*}{row}& \multirow{2}{*}{id} & \multirow{2}{*}{$y$}    & \multirow{2}{*}{$X_1$}  & \multirow{2}{*}{$X_2$}      & \multirow{2}{*}{$u$}    & \multirow{2}{*}{$t$}    & Separate   &   Schmid et al. (2020)   & Super person-period \\ 
      &      &      &                  &            &        &        &   ($X_1 ,X_2$)           &($X_1 ,X_2,u$) &($X_1 ,X_2,u,t$) \\
      \hline
      \hline
    1 & 1    & 0    & $x_{11}(0)$      & $x_{21}$   & 1      & 0      & \multirow{4}{*}{$(0,1)$} & \multirow{9}{*}{$\{(0,u): u=1,\ldots,4\}$ } & \multirow{19}{*}{\makecell{All possible \\ combinations of \\ $(t,u)$: $t<u$, \\ $t=0,1,\ldots,4$, \\ $u=1,2,\ldots,4$.}}\\ 
    \cline{1-7}
    2 & 2    & 0    & $x_{12}(0)$      & $x_{22}$   & 1      & 0      &                          &         &                  \\ 
    \cline{1-7}
    3 & 3    & 0    & $x_{13}(0)$      & $x_{23}$   & 1      & 0      &                          &             &              \\ 
    \cline{1-8}
    4 & 1    & 1    & $x_{11}(0)$      & $x_{21}$   & 2      & 0      & \multirow{3}{*}{$(0,2)$} &         &\\ 
    \cline{1-7}
    5 & 2    & 0    & $x_{12}(0)$      & $x_{22}$   & 2      & 0      &                          &       &  \\
    \cline{1-7}
    6 & 3    & 0    & $x_{13}(0)$      & $x_{23}$   & 2      & 0      &                          &        &  \\ 
    \cline{1-8}
    7 & 2    & 0    & $x_{12}(0)$      & $x_{22}$   & 3      & 0      & \multirow{2}{*}{$(0,3)$} &           & \\ 
    \cline{1-7}
    8 & 3    & 0    & $x_{13}(0)$      & $x_{23}$   & 3      & 0      &                          &          &                 \\ 
    \cline{1-8}
    9& 2    & 0    & $x_{12}(0)$      & $x_{22}$   & 4      & 0      & $(0,4)$                  &           &                \\ 
    \cline{1-9}
    10& 1    & 1    & $x_{11}(1)$      & $x_{21}$   & 2      & 1      & \multirow{3}{*}{$(1,2)$} &   \multirow{6}{*}{$\{(1,u): u=2,\ldots,4\}$ }     & \\ 
    \cline{1-7}
    11& 2    & 0    & $x_{12}(1)$      & $x_{22}$   & 2      & 1      &                          &         &                  \\ 
    \cline{1-7}
    12& 3    & 0    & $x_{13}(1)$      & $x_{23}$   & 2      & 1      &                          &         &                  \\ 
    \cline{1-8}
    13& 2    & 0    & $x_{12}(1)$      & $x_{22}$   & 3      & 1      & \multirow{2}{*}{$(1,3)$} &      &                    \\ 
    \cline{1-7}
    14& 3    & 0    & $x_{13}(1)$      & $x_{23}$   & 3      & 1      &                          &      &                    \\ 
    \cline{1-8}
    15& 2    & 0    & $x_{12}(1)$      & $x_{22}$   & 4      & 1      & $(1,4)$                  &   &                       \\ 
    \cline{1-9}
    16& 2    & 0    & $x_{12}(2)$      & $x_{22}$   & 3      & 2      & \multirow{2}{*}{$(2,3)$} &  \multirow{2}{*}{$\{(2,u):u=3,4\}$}& \\                  
    \cline{1-7}
    17& 3    & 0    & $x_{13}(2)$      & $x_{23}$   & 3      & 2      &                          &      &                    \\ 
    \cline{1-8}
    18& 2    & 0    & $x_{12}(2)$      & $x_{22}$   & 4      & 2      & $(2,4)$                  &      &                    \\ 
    \cline{1-9}
    19& 2    & 0    & $x_{12}(3)$      & $x_{22}$   & 4      & 3      & $(3,4)$                  &  $(3,4)$ &\\ 
    \hline

    \end{tabular}

    \label{allthree}
    \end{table}

Table \ref{allthree} provides an illustration of the training data set used for all three approaches to solve each of the 10 estimation problems given in Table \ref{tbl:problems}. The person-period data set is reformatted based on the generic data set given in Table \ref{generic}. Each subject has one row for each pair value of $(t,u)$ where it was still at risk of experiencing the event, i.e. its event time and censoring time both have not yet occurred at $u-1$. Only the first three subjects (up to id $=3$) are shown in the table to save space. 
For example, to solve the estimation problem for the pair $(t^\ast,u^\ast)=(1,2)$, i.e. to estimate the hazard probability for any subject at time point 2 with its covariate information at time point 1, the training data set used for the separate method would be the one given in rows 10-12 in Table \ref{allthree}. Note that only the subjects whose event time and censoring time both have not yet occurred at $u^\ast-1=1$ are included. The outcome $y$ has a value of 1 if the event occurred at time point $u^\ast=2$ and 0 otherwise. Two covariates are used for this method, $X_1$ and $X_2$. 
For the same problem, the \citet{Schmidetal2020} method uses the training data set as given in rows 10-15 in Table \ref{allthree}, and adds $u$ as a covariate. 
The third approach uses $X_1$, $X_2$, $u$ and $t$ as covariates. Its training data set consists of all of the rows of the person-period data. 
One can see that, to produce ten estimated hazard probabilities, one for each estimation problem as given in Table \ref{tbl:problems}, the first approach builds 10 models (one for each pair of $(t,u)$), the second approach builds 4 models (one for each $t$), and the third approach builds only one model (one for all pairs of $(t,u)$).

In the simulations summarized in the next section, we investigate these three approaches applied to random forest methods: separate forests, forests using the \citet{Schmidetal2020} method, and a forest built on the super person-period data set, which will be referred to as ``Separate,'' ``Poolt,'' and ``Superpp,'' respectively.

We also compare the performance of these three methods to the following two methods in the simulation study:\\
(1) Super person-period forest with baseline information only. That is, Superpp using only the covariate information at $t=0$. This method will be referred to as ``Superpp0''. \\
(2) DTPO model using the super person-period construction. This method will be referred to as ``SuperppDTPO''. 

Note that SuperppDTPO targets the log-linear survival relationship, and Superpp0 is a non-parametric method, but never updates the information from the initial status. These two methods serve as benchmark parametric and nonparametric methods, respectively, as we investigate the performance of the three methods under different model setups.

\begin{figure}[!t]
    \centering
    \includegraphics[scale = 0.51]{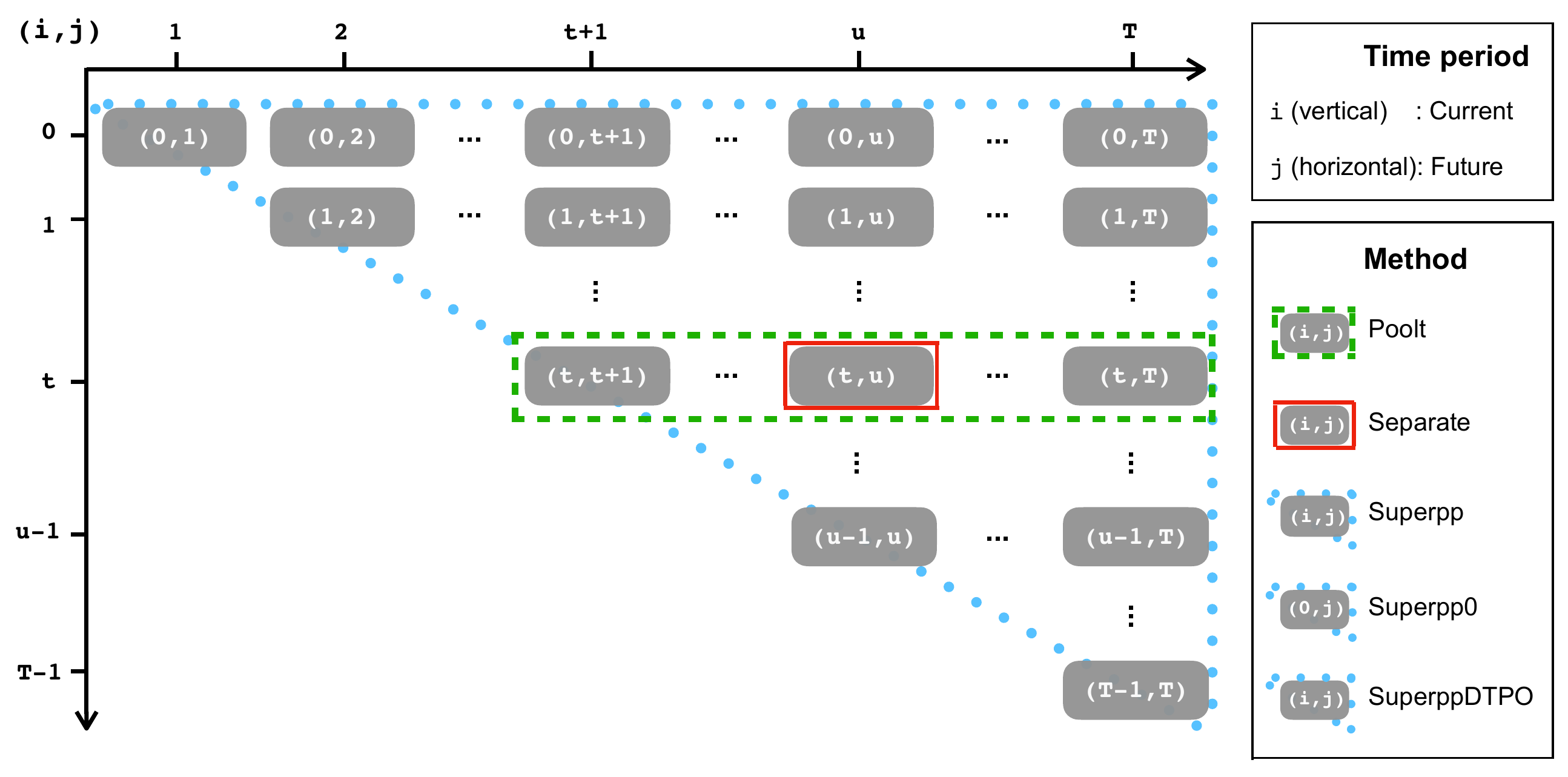}
    \caption{Graphical overview of the methods for dynamic estimation. } \label{fig:graphic_methods}
\end{figure}

Figure \ref{fig:graphic_methods} provides a graphical overview of the methods for dynamic estimation. Consider the set of time points $\{0, 1,\ldots, T\}$. Each entry $(i, j)$ contains the subjects that are still alive and not censored at time $j-1$, and it gives the information available at time $i$ from each subject in that cell. 
Suppose we are at the current time point $t$ and want to estimate the hazard function for some future time point $u (> t)$. The red box (solid line) contains all of the subject information that Separate uses for training the forest model, the green box (dashed line) contains all that Poolt uses, and the blue triangular region (dotted line) contains all that Superpp uses. Note that SuperppDTPO uses the same subject information as Superpp does, and Superpp0 also uses the same subjects but with all $(i,j)$ replaced by $(0,j)$ indicating it uses the baseline information only. There is in total one blue region, $T$ green boxes, and $T \times (T + 1) / 2$ red boxes, implying the construction of one Superpp forest, one Superpp0 forest, one SuperppDTPO model, $T$ Poolt forests, and $T \times (T + 1)/2$ Separate forests are used to construct estimates of hazards for all combinations of $(t,u)$ for a given value of $T$. Note that Table \ref{allthree} is a specific example of this construction where there are only three subjects and $T=4$.

\section{Simulation Study}\label{sec:simulations}
\texttt{R} \citep{R2020} was used to perform the simulations. The package \texttt{ranger} \citep{Wrightetal2020} was used to build the forests with the Hellinger splitting rule for methods Separate, Poolt, Superpp and Superpp0, that is, all methods that require a classification forest. The number of trees in all forests is $500$. SuperppDTPO was implemented using logistic regression on the Superpp data set.

\subsection{Simulation Design}\label{sec:simulations_design}
The data generating process (DGP) is a discretized version of the continuous-time survival data generated from the model used in the simulation study in \citet{Yaoetal2020}. We consider the following factors for different variations of DGPs. 

\noindent 
(1) Different combinations of numbers of time-invariant and time-varying covariates in the true generating model (Scenario). 

\noindent 
(2) Different matrices to generate covariates' values with autocorrelation for the time-varying variables (labelled as ``Strong'' and ``Weak''). Note that stronger autocorrelation would imply that covariate values from earlier time points would tend to be more similar to those in later time points, making future estimation easier.

\noindent 
(3) Different signal-to-noise ratios (SNR) labelled as ``High'' and ``Low,'' constructed by choosing different coefficients in the model. 

\noindent 
(4) Different survival distributions: Exponential, Weibull and Gompertz. 

\noindent 
(5) Different survival relationships between the hazards and covariates: a log-linear one, a log-nonlinear one and a log-interaction model. 

\noindent 
(6) Different censoring rates: 10\% and 50\%.

\noindent 
(7) Different training sample sizes: $n=200,1000$ and $5000$.

\noindent
(8) Different total numbers of time points: $T=4$ and $8$.

Each model is fit with a training sample of size $1000$. The performance of the fitted models is then evaluated with $T$ independent test sets of size 1000 each. The $k$-th test set ($k=1,2,\ldots,T$) includes only the subjects that are still at risk at $u=k$, so it can be used when $(t,u)=(j,k)$ for all $j = 0,1,\ldots,k-1$. 
Each simulation is repeated 500 times.
See Section S1.1 in the supplemental material for more details of the simulation design.

\subsection{Simulation Results}
We consider three criteria for evaluating the accuracy of the methods: the absolute  difference (ADIST), the absolute log odds ratio (ALOR) and the concordance index (C-index) for hazard rates. Let $\hat{h}$ and $h$ be the estimated and the true hazards. ADIST is defined by
\begin{align*}
   \mathrm{ADIST}\big(h, \hat{h}\big) = |\hat{h}-h|,
\end{align*}
and ALOR by 
\begin{align*}
    \mathrm{ALOR}\big(h, \hat{h}\big) = \vert \ln ((\hat{h} (1-h))/((1-\hat{h})h)) \vert. 
\end{align*}
Both ADIST and ALOR take a minimum value of 0 when $\hat{h}=h$ while ALOR also takes the magnitude of $h$ and $\hat{h}$ into account. The C-Index computes the proportion of concordant pairs over all possible evaluation pairs:
\begin{align*}
    C=\frac{\sum_{i\neq j}\mathbb{I}\big(h_i>h_j\big)\cdot\mathbb{I}\big(\hat{h}_i>\hat{h}_j\big)}{\sum_{i\neq j}\mathbb{I}\big(h_i>h_j\big)},
\end{align*}
where the indices $i$ and $j$ refer to pairs of hazards in the test sample for a given combination of $(t, u)$. It is designed to estimate the concordance probability $\mathbb{P}\big(\hat{h}_i>\hat{h}_i\mid h_i>h_j\big)$, which compares the rankings of two independent pairs of hazard rates $h_i$, $h_j$ and estimates $\hat{h}_i$, $\hat{h}_j$. The concordance probability evaluates whether values of $\hat{h}_i$ are directly associated with values of $h_i$. 
Note that while both ADIST and ALOR measure the distance between the true hazard and its estimate, the C-index is a rank-based metric that evaluates if the true and estimated values are ordered similarly, and a high value does not necessarily imply that the estimated values are close to the true ones. 

Extensive simulation studies show that the total number of time points $T$ in the true model does not affect the general conclusions. In the following discussion, we focus on the cases where $T=4$ (see Table S1.2 in the supplemental material for performance comparison between $T=4$ and $T=8$).

\begin{figure}[t]
    \centering
    \includegraphics[scale = 0.55]{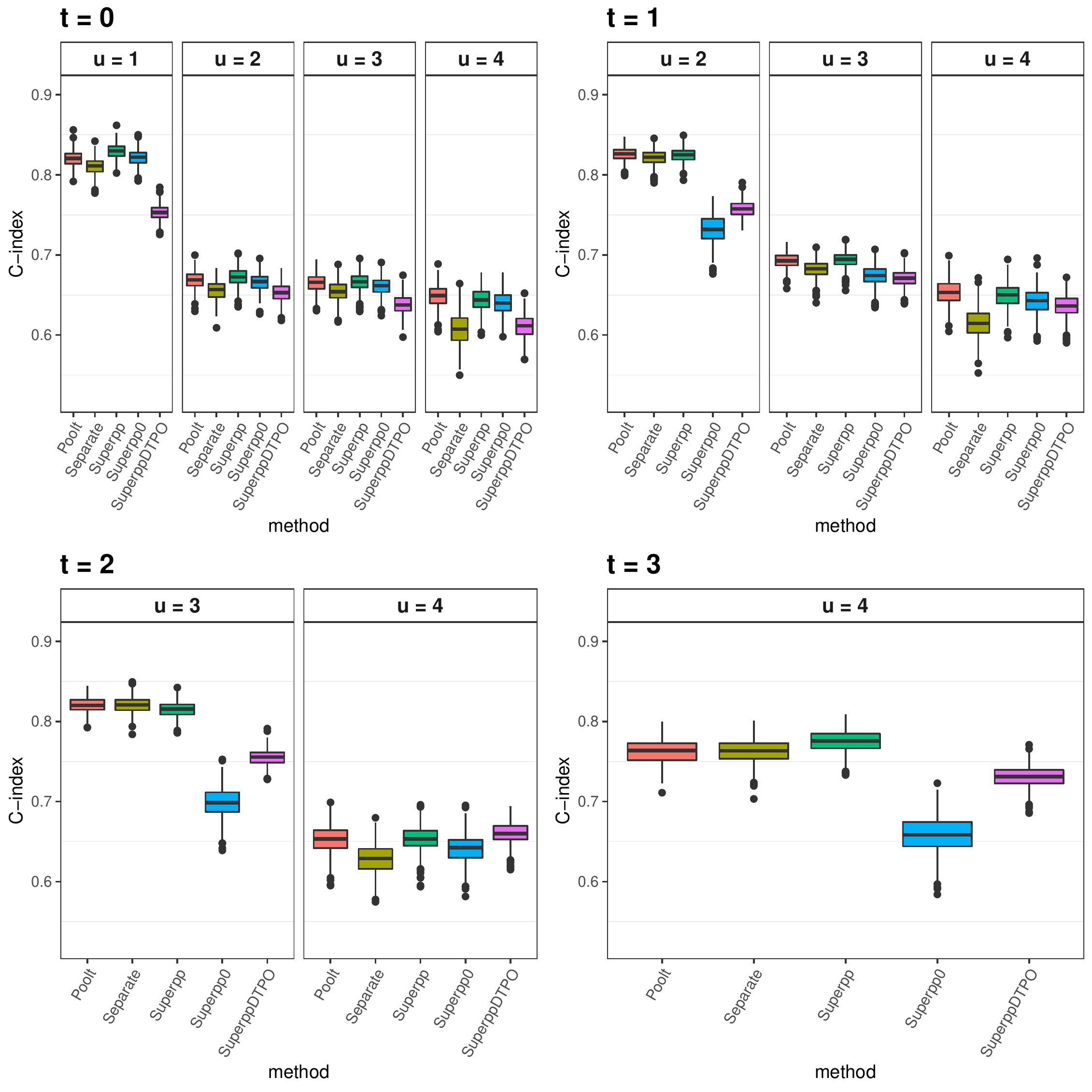}
    \caption{Simulation results comparing the distribution of C-index on test sets across methods for each pair of $(t,u)$, trained on sample data of size 1000, with 10\% censoring rate, generated following a Weibull distribution with an interaction survival relationship in the scenario 2TI $+$ 4TV, with high SNR and strong autocorrelation. The total number of time points is $T = 4$.}
    \label{fig:cindex_example}
\end{figure}

Boxplots for the 500 simulation runs of each method for each combination $(t, u)$ based on the evaluation of ADIST and C-index are provided in Section S1.2 in the supplemental material. Boxplot results from ALOR are not reported since the conclusions are essentially the same as those from ADIST (ALOR results for performance comparison are still provided in summary tables in Section S1.3 in the supplemental material). Figures \ref{fig:adist_example} and \ref{fig:cindex_example} give an example of the boxplots for ADIST and C-index, respectively, when the training sample size is 1000, the censoring rate is 10\%, and the data are generated following a Weibull distribution with an interaction survival relationship in the scenario 2TI $+$ 4TV (2 time-invariant and 4 time-varying covariates), with high SNR and strong autocorrelation, and with a total number of time points of $T=4$. In general, for a given $t$ (i.e. for a given plot), the performance of the methods usually worsens as $u$ increases. This is expected, since it is more difficult to estimate the hazard for horizons further away. 
\begin{figure}[t]
    \centering
    \includegraphics[scale = 0.55]{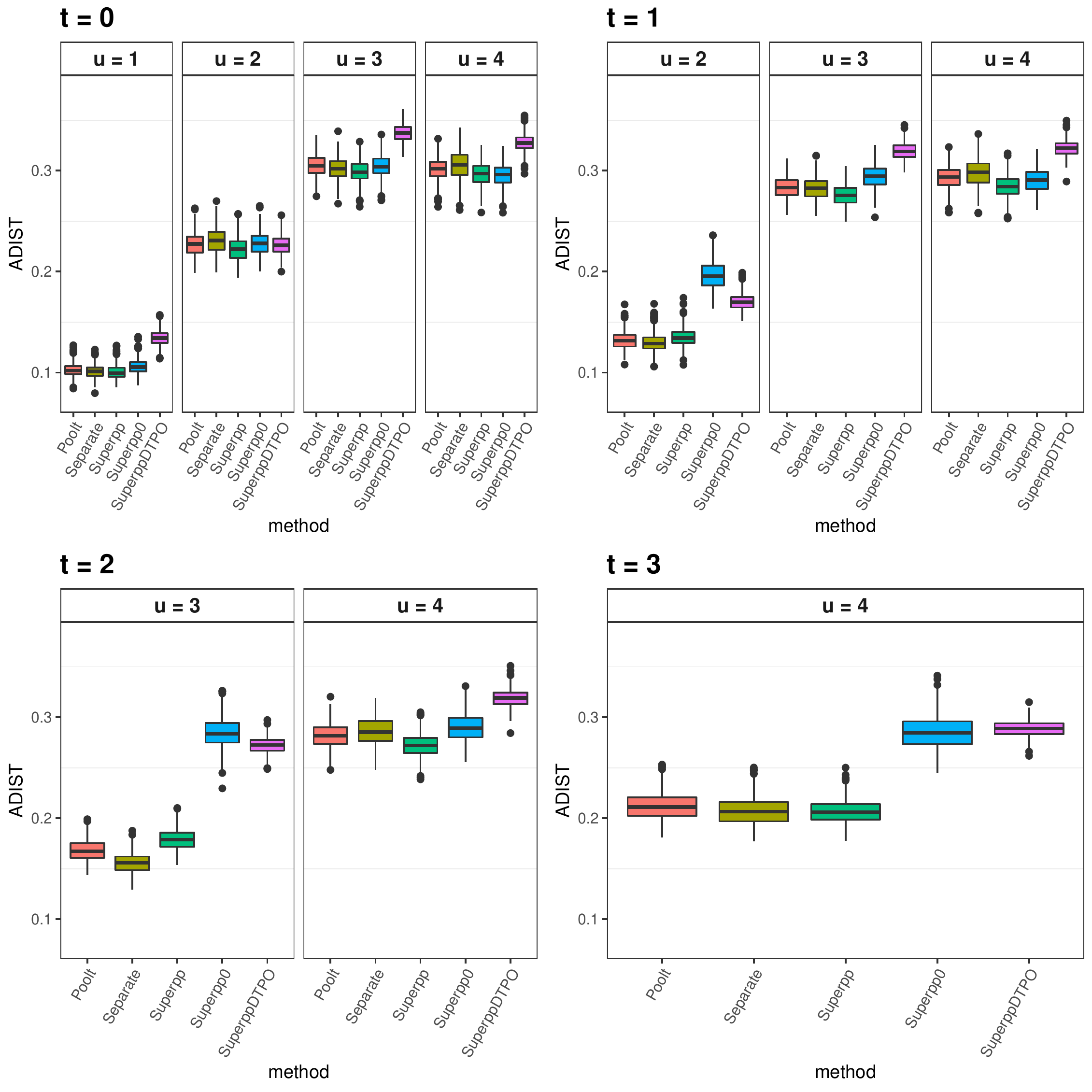}
    \caption{Simulation results comparing the distribution of ADIST on test sets across methods for each pair of $(t,u)$, trained on sample data of size 1000, with 10\% censoring rate, generated following a Weibull distribution with an interaction survival relationship in the scenario 2TI $+$ 4TV, with high SNR and strong autocorrelation. The total number of time points is $T = 4$.}
    \label{fig:adist_example}
\end{figure}

From the boxplots based on ADIST evaluation, the parametric SuperppDTPO method works well as expected when the underlying survival relationship is linear. In most of the other cases, it is outperformed by the non-parametric forest methods. 
Superpp always gives a better performance for dynamic estimation than Superpp0 , which is again expected as the latter only uses the baseline covariates' values. In general, the three forest methods that use all of the covariates' information, Separate, Poolt and Superpp, perform the best compared to the other two simpler methods, which presumably reflects that the hazard estimates from the three forests are less biased in general due to the flexibility of the estimators.

Note that the boxplot results for evaluation from ADIST and those from C-index do not always agree with each other. In particular, the C-index tends to favor SuperppDTPO in general. For example, Figure \ref{fig:cindex_example} shows that SuperppDTPO outperforms Separate when $(t,u) = (1,4)$ and dominates the other methods when $(t,u) = (2,4)$ while in Figure \ref{fig:adist_example} it gives the worse performance among all methods in both cases. As noted, ADIST is a calibration metric while C-index is a rank-based metric. Bias is more important for accurate estimation of hazards, while variance is more important for accurate ordering of hazards. This results in favorable performance for forests using the time-varying information for the ADIST criterion, and sometimes a favorable performance for the parametric and the simpler forest that uses only the baseline information for the C-index criterion.

We now focus on the three forest methods, Separate, Poolt and Superpp. Summary tables that provide the ranking of these three methods for performance comparison using ADIST, ALOR and C-index for each factor separately are given in Section S1.3 in the supplemental material. In each situation, the Poolt method always ranks between Separate and Superpp, so we focus on comparison between Separate and Superpp. Specifically, the comparison is carried out under two situations separately, when the estimation horizon $(u-t)$ is equal to $1$ and when it is larger than $1$. We give $T=4$ as an example. In each situation, using factorial designs, we study the difference of a given measure between Separate and Superpp under the effects of the following factors: autocorrelation, censoring rate, survival distribution, survival relationship, training sample size, scenario and SNR. 
The effects are estimated based on an analysis of variance model fit with these factors as main effects.

\begin{figure}[t]
    \centering
    \includegraphics[scale = 0.6]{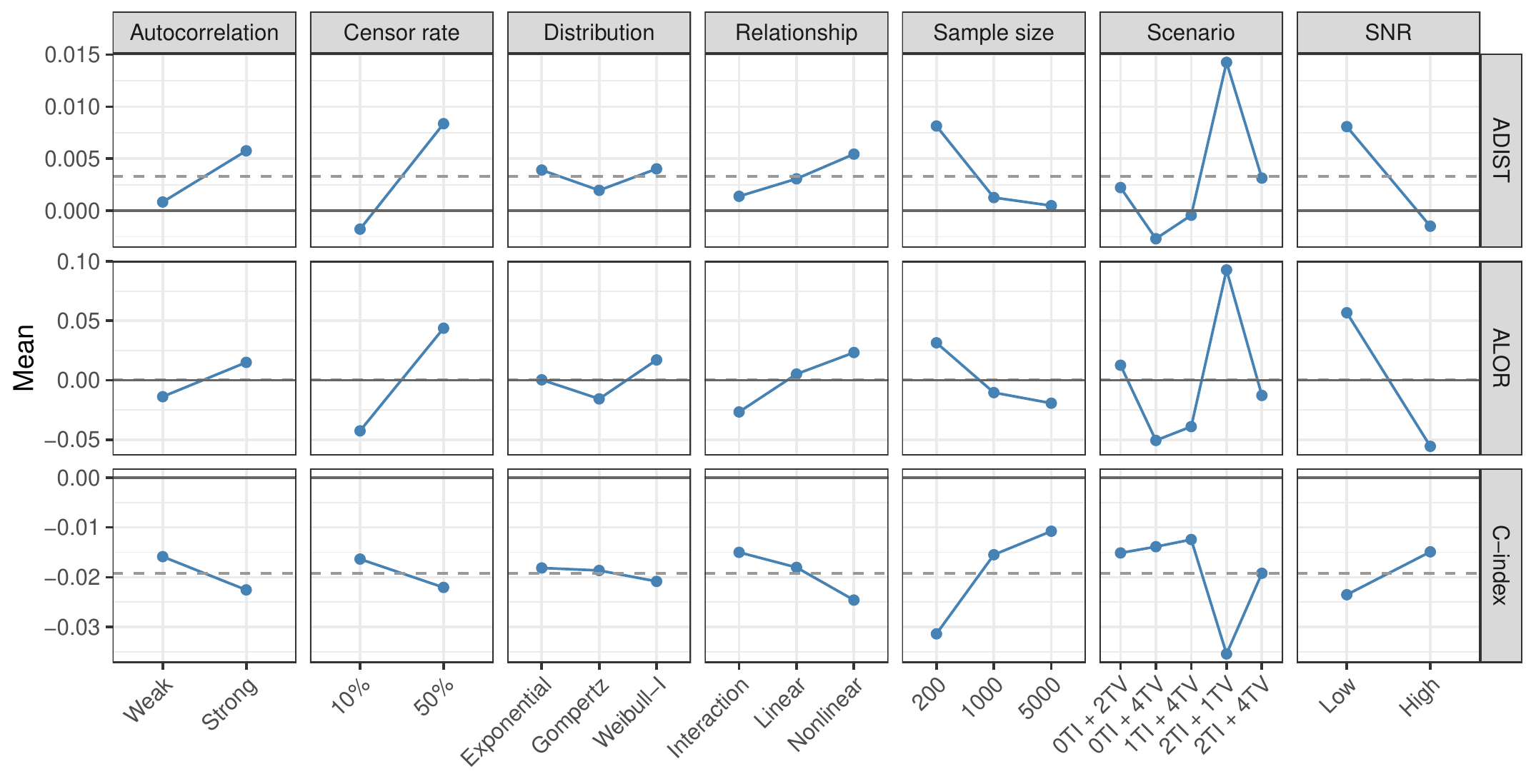}
    \caption{Main effect plot of difference for each measurement between the Separate and the Superpp method for estimation horizon $(u-t)=1$, that is, one-step ahead estimation, when $T=4$. Given any measurement $m$, the difference is computed as $m_{\mathrm{Separate}} - m_{\mathrm{Superpp}}$. The solid line gives the zero value and the dashed line gives the mean value over all effects for reference.}
    \label{fig:maineffects_horizon3equals1}
\end{figure}

\begin{figure}[t]
    \centering
    \includegraphics[scale = 0.6]{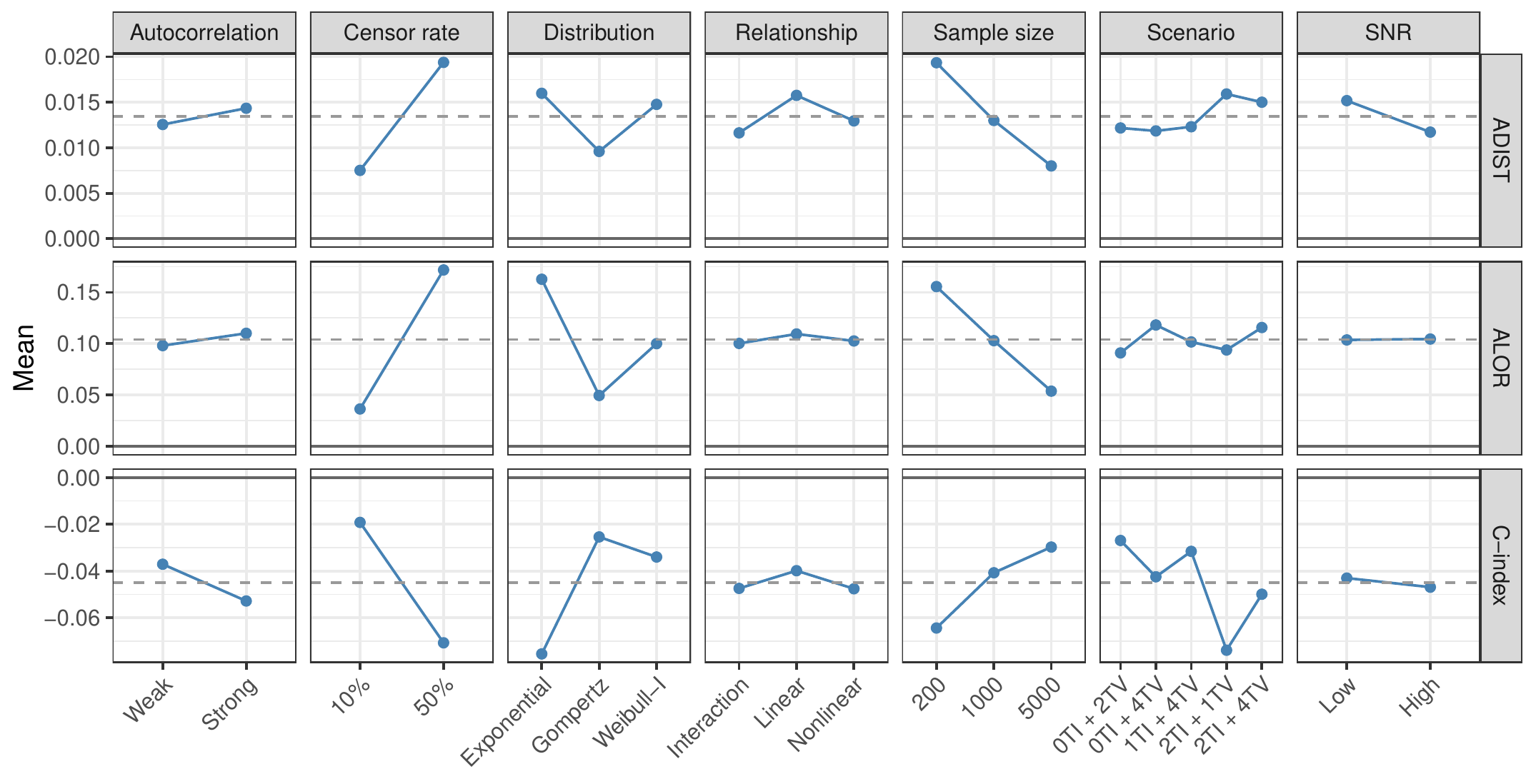}
    \caption{Main effect plot of difference for each measurement between the Separate and the Superpp method for estimation horizon $(u-t)>1$, that is, more than one-step ahead estimation, when $T=4$. Given any measurement $m$, the difference is computed as $m_{\mathrm{Separate}} - m_{\mathrm{Superpp}}$. The solid line gives the zero value and the dashed line gives the mean value over all effects for reference.}
    \label{fig:maineffects_horizon3equals0}
\end{figure}

Figures \ref{fig:maineffects_horizon3equals1} and \ref{fig:maineffects_horizon3equals0} provide main effects plots for the difference between Separate and Superpp under all three measurements for $(u-t) = 1$ and $(u-t) > 1$, respectively. In both cases, for each given effect, the general pattern of the change in difference by varying the level of the effect is the same for ADIST and ALOR, and opposite for the C-index (recall that low values of ADIST and ALOR and high values of C-index reflect better comparative performance of Separate over Superpp). Superpp is always the best performer for $(u-t)>1$, although the effects are weaker than for $(u-t)=1$, reflecting the difficulties of predicting farther in the future. We therefore focus on estimation horizon $(u - t)= 1$ in the following discussion. 

We first examine the results based on ADIST. The overall center of location is positive, highlighting that Superpp performs generally better than Separate. However, Separate can improve relative to Superpp with changes in factors. The larger the training sample size, the higher the SNR, or the smaller the censoring rate, the stronger the ability of any method to estimate the underlying survival relationship. In that situation, the flexibility of the Separate method is advantageous, while the stability of pooling is advantageous when the underlying relationship is more difficult to estimate. It is clear that the difference between the number of time-invariant (TI) and the number of time-varying (TV) covariates is driving the scenario effect. When \#TI $-$ \#TV $= 1$, Superpp is the big winner; when \#TI $-$ \#TV $= -2$, Superpp still wins, but by a smaller margin; when \#TI $-$ \#TV $= -3$, Separate wins; and when \#TI $-$ \#TV $= -4$, Separate wins by the largest margin. Presumably, this reflects that the Separate method is more sensitive to local time-varying effects, while pooling benefits from the stability associated with covariates being time-invariant.

Separate performs better relative to Supperpp when using ALOR as the measure of accuracy (sometimes beating it), reflecting that it can estimate extreme hazards more effectively. This is caused by the pooling underlying Superpp shrinking the estimated hazards away from the extremes; see the corresponding plot and discussion in Section S1.4 of the Supplemental material.

The relative performance of Separate and Superpp using C-index is similar to that using ADIST, with Superpp being most effective. This may be explained by the fact that pooling reduces the variance and thus makes Superpp superior when we evaluate the performance with C-index.

Overall, weaker autocorrelation in covariates, higher censoring rate, smaller training sample size, smaller portion of covariates being time-varying, lower SNR, and estimation further in the future all reflect more difficult estimation tasks, and the less flexible but more stable pooling approach dominates. Conversely, in the opposite situations where signals are stronger and noise less extreme, the more flexible but more variable Separate approach is more effective.

\section{Concluding Remarks}\label{sec:conclusions}
This paper provides an investigation of different discrete-time survival forest methods for dynamic estimation with time-varying covariates. All methods investigated can be easily implemented using existing \emph{R} packages. The results show that all methods perform well and none dominates the others. As a general rule, situations that are more difficult from an estimation point of view (such as weaker signals and less data) favor a global fit, pooling over all time points, taking advantage of reduced variance, while situations that are easier from an estimation point of view (such as stronger signals and more data) favor local fits, taking advantage of increased flexibility.

It should be noted that the methods discussed here all assume that censoring is uninformative; that is, subjects are censored for reasons unrelated to the time to event being examined. This is potentially an issue in the bankruptcy data examined in Section S2 in the supplemental material, as it is possible that companies that are in danger of declaring bankruptcy stop filing financial disclosures in order to hide their precarious financial position. A common parametric approach to this problem is the use of joint modeling, in which the assumed parametric forms for longitudinal predictors and the time to event are linked through shared random effects \citep{Rizopoulos2012}. It is possible that such models could be generalized to the discrete survival situation to allow tree-based structures on the joint distribution, perhaps based on recently-developed tree-based methods for longitudinal data such as those described in \citet{Hajjemetal2011, Hajjemetal2014}, \citet{SelaandSimonoff2012}, and \citet{FuandSimonoff2015}.

In this paper we have limited ourselves to an event that is only incomplete due to right-censoring. Other reasons that the actual time to event is hidden are possible, such as left-truncation and interval censoring. Generalization of the methods discussed here would be useful future work, perhaps based on the tree- and forest-based methods for continuous time-to-event data discussed in \citet{FuandSimonoff2017a,FuandSimonoff2017b} and \citet{Yaoetal2021}.

Presumably, all information of the time-varying covariates is available up to the given time for the estimation of the hazard function at a future time point. In this paper, we implement the forest methods based only on the current (latest) values of the time-varying covariates without including any lagged values. Future work can be done to investigate how to efficiently use the available lags, including the associated variable selection problems.

\section*{CONTENTS OF SUPPLEMENTAL MATERIAL}
A separate supplemental material document provides more details about the simulation study and a real data example using bankruptcy data. It consists of the following sections: 
\begin{enumerate}[label = S\arabic*]
    \item Simulation study
    \begin{enumerate}
        \item[S1.1] Data generation processes
        \item[S1.2] Box-plots for each combination of $(t,u)$
        \item[S1.3] Summary tables 
        \item[S1.4] The shrinkage effect of the Superpp method
    \end{enumerate}
    \item Real data example
\end{enumerate}

In addition, the datasets generated and analysed in the simulation study are available from the github repository,\\
\textit{https://github.com/ElainaYao/DynamicEstimationDTSD}, including R scripts for reproducibility of the simulations.

\section*{ACKNOWLEDGEMENTS}
We would like to thank the associate editor and three anonymous reviewers for their interesting and constructive comments that led to an improved version of this paper. D Larocque acknowledges the financial support of The Natural Sciences and Engineering Research Council of Canada (NSERC) and Fondation HEC Montr\'{e}al.

\bibliographystyle{chicago}
\bibliography{reference.bib}

\end{document}